\documentclass[10pt]{iopart}

\usepackage[utf8]{inputenc}
\usepackage{graphicx}
\usepackage{cite}

\usepackage{iopams}  

\usepackage[T1]{fontenc}

\usepackage[switch]{lineno} 
\usepackage{lipsum} 

\usepackage{array}   

\begin{document}


\title{Validity of point-mass model in off-resonance dynamic atomic force microscopy}

\author{Shatruhan Singh Rajput$^1$, Surya Pratap S Deopa$^1$, V. J. Ajith$^1$, Sukrut C. Kamerkar$^2$, and Shivprasad Patil$^{1*}$}

\address{$^1$ Department of Physics, Indian Institute of Science Education and Research Pune, Dr. Homi Bhabha Road, Pashan,  Pune 411008, India \newline $^2$ Department of Biology, Indian Institute of Science Education and Research Pune, Dr. Homi Bhabha Road, Pashan,  Pune 411008, India}

\ead{s.patil@iiserpune.ac.in}
\vspace{10pt}

\begin{abstract}
The quantitative measurement of viscoelasticity of nano-scale entities is an important goal of nanotechnology research and  there is  considerable progress with advent of dynamic Atomic Force Microscopy.  The hydrodynamics of cantilever, the force sensor in AFM measurements,  plays a pivotal role in quantitative estimates of nano-scale viscoelasticity. The point-mass model, wherein the AFM cantilever is approximated as a point mass with mass-less spring is widely used in dynamic AFM analysis and its validity, particularly in liquid environments, is debated. It is suggested that the cantilever must be treated as a continuous rectangular beam to obtain accurate estimates of nano-scale viscoelasticity of materials it is probing. Here, we derived equations, which relate stiffness and damping coefficient of the material under investigation to measured parameters, by approximating cantilever as a point mass and also considering the full geometric details. These equations are derived for both tip-excited as well as base excited cantilevers. We have performed off-resonance dynamic atomic force spectroscopy on a single protein molecule  to investigate the validity of widely used point-mass model. We performed measurements with AFMs equipped with  different cantilever excitation methods  as well as detection schemes to measure cantilever response. The data was analyzed using both, continuous-beam model  and the  point-mass model. We found that both models yield  same results when the experiments are performed in truly off-resonance regime with small amplitudes and the cantilever stiffness is much higher than the interaction stiffness. Our findings suggest that a simple point-mass approximation based model is adequate to describe the dynamics, provided care is taken while performing  experiments so that the approximations used in these models are valid.     
\end{abstract}

\vspace{2pc}

\ioptwocol

\section{Introduction}
The atomic force microscope (AFM) is a powerful tool to obtain the surface topography of insulators. Shortly after its invention, it  successfully produced images  of substrates with atomic resolution \cite{binnig1987atomic, alexander1989atomic, marti1988atomic}. Over time, it has been used to quantify the inter-atomic forces between the cantilever-tip and substrate atoms in UHV \cite{pethica1987tip, jarvis1996direct, jarvis1999off, jeffery2000quantitative, hoffmann2001energy, hoffmann2001direct, ozer2007local} and ambient conditions \cite{garcia1999attractive}. In the past few decades, the AFM has been recognised as a versatile tool to measure  the interaction forces between different biomolecules and forces required to unfold proteins \cite{ludwig1999afm, sotomayor2007single, hinterdorfer2006detection, dufrene2008recent, koti2008single, kotamarthi2013single}. In addition, many groups have attempted to measure the viscoelastic properties of single-molecules directly via dynamic atomic force spectroscopy  \cite{rajput2020nano, khatri2008internal, higgins2006frequency, benedetti2016can, liang2019investigating, humphris2000active, kawakami2004viscoelastic, kawakami2005viscoelastic, kawakami2006viscoelastic, bippes2006direct, khatri2007entropy, janovjak2005molecular, chtcheglova2004force, kienberger2000static}.

In a dynamic AFM experiment, an oscillating tip mounted on a cantilever force sensor is allowed to interact with a sample which causes change in parameters such as amplitude and phase. The  dynamic spectroscopy has been used extensively in UHV to study inter-atomic forces \cite{jarvis1996direct, hoffmann2001energy, hoffmann2001direct}. It is also  implemented in viscous medium to study behaviour of liquids under nano-confinement \cite{jeffery2004direct, maali2006oscillatory, patil2006solid, khan2010dynamic, de2010dissipation, khan2015squeeze, o1994atomic, kaggwa2008artifact} and single-molecules under force \cite{khatri2008internal, higgins2006frequency, liang2019investigating, humphris2000active, kawakami2004viscoelastic, kawakami2005viscoelastic, kawakami2006viscoelastic, bippes2006direct, khatri2007entropy, janovjak2005molecular, chtcheglova2004force, kienberger2000static}. 
In order to extract mechanical properties of a single molecule or nano-scale system probed by the tip, a mathematical treatment of cantilever dynamics in various media such as  Ultra High Vacuum (UHV) or liquid environment  is essential. There is a debate among AFM users about correctness of various approximations used while trying to describe the cantilever dynamics. In situations wherein the damping provided by the viscous medium to the cantilever is low, such as ambient conditions or UHV, the Point-Mass (PM) model, in which the cantilever is approximated as a point-mass attached to a massless spring,  has been used successfully for quantitative analysis of the dynamic AFM data \cite{garcia1999attractive, garcia2002dynamic}. However, for cantilever hydrodynamics in liquid environment,  validity of these approximations in various regimes have been discussed in many works \cite{raman2008cantilever, xu2007comparative, benedetti2016can, turner1997high}.  

In another approach  to describe the hydrodynamics, the  cantilever is modelled as Continuous Beam (CB) \cite{garcia2011amplitude, rabe2006atomic, kiracofe2011quantitative, butt1995calculation, sader1998frequency, lozano2009theory}. This approach gains importance particularly in liquid environments with significant viscous damping and it works with less number of approximations compared to the PM model.  Recently, the CB model  is used for base-excited, low frequency  measurements of protein pulling performed using AFM  \cite{benedetti2016can}. In this work, in order to describe the dithering cantilever-molecule system, the Euler-Bernoulli equation with appropriate boundary conditions was solved. They proposed that interaction-force alters the boundary conditions of the cantilever-molecule system and were considered explicitly in their calculations.
It is  argued  that ignoring  details of cantilever geometry and not including the interaction in boundary conditions  results in ambiguous interpretation of phase signal as dissipation from single molecule. It was suggested that the CB model is an appropriate model to extract viscoelasticity of nano-scale interactions and in-phase ($X$) and out-of-phase ($Y$) components of the oscillations are observable parameters, which are directly related to stiffness and damping.

In point-mass model, cantilever-beam is approximated as a point-mass and the interaction force acting on the cantilever-tip is treated as a small perturbation on the moving point-mass. These assumptions make the understanding of cantilever dynamics simple and its analytical solution becomes straight-forward. This model has been extensively used in the past due to its simplicity. The use of CB to model the cantilever raises doubt on previous works  that have been done on various nano-scale systems such as single polymer \cite{liang2019investigating, kawakami2006viscoelastic, janovjak2005molecular, khatri2008internal} and nano-confined liquids \cite{de2010dissipation, maali2006oscillatory}. These experiments have used  base-excited cantilevers with deflection detection type sensing and  PM model was used for data analysis. 
Apart from this, a large body of work  exists to understand the behaviour of nano-confined liquids using base-excited dynamic measurements using interferometer based AFM in which the tip  displacement is measured  and PM model has been used for data analysis \cite{khan2010dynamic, patil2006solid, khan2015squeeze, jeffery2004direct}.
The reliability of these  results  depend on the correctness  of the mathematical model  that has been used to analyse the data. A judicious choice of  model which  correctly captures cantilever hydrodynamics  for a particular experimental strategy is of utmost importance before conclusions about nano-scale dissipation in liquid environments are drawn. 


In this article, we have treated the cantilever dynamics in liquid environments in two ways,  i) the geometric details of the cantilever as a continuous beam are incorporated in Euler-Bernolli eqution of motion and  ii) the cantilever is approximated as a point-mass with a mass-less spring whose stiffness is determined by the first eigen mode of the continuous beam. Using these two approaches we derived equations to relate the stiffness and damping coefficient of the nanoscale entity under consideration to the experimentally measured quantities such as phase and amplitude of the oscillating cantilever. We developed solutions for both tip-excited as well as based excited cantilevers. Furthermore, we have investigated the validity of PM model to analyse the dynamic AFM measurements performed in liquid environments.

We compare the solutions of these two models within  specific experimental strategies used to excite as well as detect the cantilever response.  We have performed two types of measurements. i) Base-excited cantilever with interferometer based detection scheme wherein cantilever tip displacement is measured. ii) Tip-excited cantilever with deflection detection scheme which measures the bending (or slope) of the cantilever at the tip end. The experimental data is analysed using both PM and CB models to check the validity of PM in these measurement schemes. Further, the requirements on experimental scheme in order to use PM model to extract the stiffness and dissipation coefficient of single protein molecules are deduced.  We find that PM model is successful in estimating quantitative viscoelastic response  of nano-scale  entities if care is taken while exciting the  cantilever  or detecting its response. In general, true  off-resonance operation, stiff cantilevers and small amplitudes are key elements in successful quantification of viscoelasticity using PM model in amplitude modulation dynamic AFM.

\section{Theory}

\textbf{{Equation of motion:}} \\
Bending of a homogeneous rectangular cantilever beam in a viscous medium is described by Euler-Bernoulli equation. The internal damping of the cantilever beam has been ignored. 

   \begin{equation} 
        -\tilde{\rho}\tilde{S}\frac{\partial^2 y(x,t)}
        {{\partial t^2 }} - \gamma_c \frac{\partial y(x,t)}{\partial t} = EI \frac{\partial^4 y(x,t)}{\partial x^4}
        \label{E-Euler-Bernoulli-equation}
    \end{equation}

Where $x$ is the coordinate along the cantilever length with $x=0$ at the clamped end and $x=L$ at the free end. $y$ is the displacement along the perpendicular direction of the cantilever length. $\tilde{\rho}\tilde{S} = \rho S + m_a$, $m_a$ is hydrodynamic added mass where $\rho$ is mass density of the cantilever material, $S(=bh)$ is area of the cantilever cross-section perpendicular to its length, $b$ and $h$ are width and thickness of the cantilever respectively, $\gamma_c$ is the cantilever drag coefficient per unit length, $E$ is Young's modulus and $I(=bh{^2}/12)$ is second area moment.

Analytical solution for Eq. \ref{E-Euler-Bernoulli-equation} can be derived for few limiting cases in two ways. One method accounts for the continuous nature of the cantilever-beam in derivation and we refer to it  as  \textit{continuous-beam (CB) model}. Another method assumes to be the cantilever a point-mass and it is known as \textit{point-mass (PM) model}. In CB model, the solution of Eq. \ref{E-Euler-Bernoulli-equation} is assumed to be $y(x,t) = y(x) \:e^{i\omega t}$, whereas, to arrive at PM model, method of variable separation is  used first where solution is assumed to be of the form $y(x,t) = y(x) Y(t)$. $y(x)$ is the general solution of space part of Eq. \ref{E-Euler-Bernoulli-equation}, which is required to be solved in both models. Space part of Eq. \ref{E-Euler-Bernoulli-equation} is,  

    \begin{equation} 
        \frac{d^4 y(x)}{dx^4} =k^4 y(x)
        \label{S_spatial_part}
    \end{equation}
    
\noindent and its general solution is:

    \begin{equation*} 
        y(x) = a\sin{(kx)} + b\cos{(kx)} + c\sinh{(kx)} + 
    \end{equation*}
    \begin{equation} \label{E-gen_sol_PM_derivation}
        \qquad \qquad \qquad \qquad \qquad \qquad \quad  d\cosh{(kx)}
    \end{equation}

\noindent Where $a$, $b$, $c$, and $d$ are constants to be determined using four appropriate boundary conditions. These boundary conditions are derivatives of $y(x)$ at some values of $x$.  They are determined by experimental situations such as excitation method as well as the interaction forces experienced by the tip.  In typical AFM experiments, interaction force acts at the tip end of  the cantilever and excitation can either be applied at the base or the tip. 

The two models differ in consideration of boundary conditions. In CB model, cantilever excitation and interaction forces are accounted for in the boundary conditions. Whereas, in PM model, a set of homogeneous boundary conditions are applied and cantilever excitation and interaction force are included in the equation of motion.

\subsection{\textbf{Continuous-beam (CB) model}}

In CB model, continuous-nature of the cantilever beam is considered. Solution of Eq. \ref{E-Euler-Bernoulli-equation} is assumed to be $y(x,t) = y(x) \:e^{i\omega t}$. Substituting the solution in Eq. \ref{E-Euler-Bernoulli-equation} results in Eq. \ref{S_spatial_part} (space-part) and following the dispersion relation:

   \begin{equation} 
        \tilde{\rho}\tilde{S}\omega^2 - i\omega\gamma_c = EIk^4
        \label{E-z_realtion}
    \end{equation}

To solve Eq. \ref{E-Euler-Bernoulli-equation}, cantilever excitation and interaction force are included in boundary conditions. Interaction force is considered as linear ($F_i = k_i\:y(t) + \bar{\gamma_i}\:\frac{dy(t)}{dt}$) and solved for small-oscillation frequency ($z<<1$ or $\omega << \omega_0$) and small interaction force ($g<<1$ or $k_i << k_c$). $k_c$ and $\omega_0$ are stiffness and resonance frequency of the cantilever respectively. Parameters $z$ and $g$ are defined as follows:

       \begin{equation*}        
            z = kL = \Big[\frac{3(\tilde{\rho}\tilde{S} \omega^2 L - i\omega\gamma_c L)}{k_c}\Big]^{\frac{1}{4}}, 
      \end{equation*}
      \begin{equation}  \label{E-assumptions}
            g = \frac{3(k_i+i\omega\bar{\gamma}_i)}{k_c}
      \end{equation}

To get the complete solution of Eq. \ref{E-Euler-Bernoulli-equation} for base and tip excitation, Eq. \ref{S_spatial_part} is required to be solved separately. Subscripts $'b'$ and $'d'$ stand for bending and displacement respectively in the entire manuscript. \\

\begin{figure}[hbt!]
  \centering
  \includegraphics[width=1\linewidth]{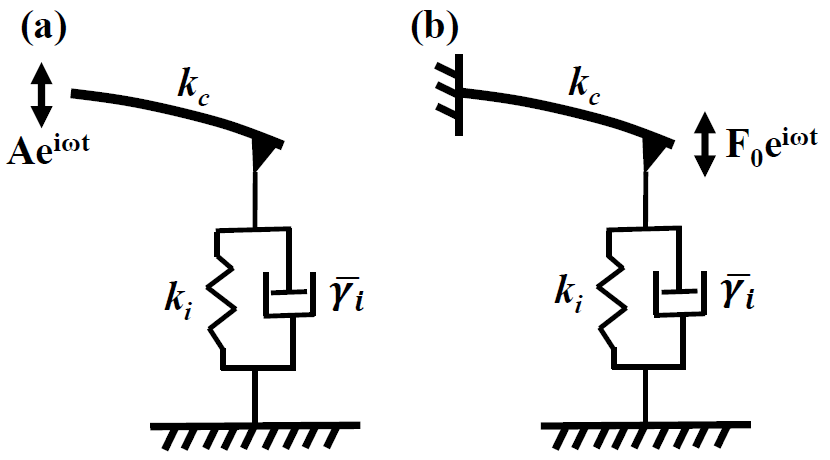}
  \caption {Schematic representation of oscillating cantilever  under influence of a linear interaction force. In off-resonance operation the  medium damping is negligibly small and hence not shown in the schematic. (a) Representing the base-excited cantilever where base is driven by $Ae^{i\omega t}$. (b) Representing the tip-excited cantilever where base is rigidly fixed and tip is excited with force $F_0e^{i\omega t}$.}
 \label{F-CM_base-tip_excitation_schematic}
 \end{figure}

\subsubsection{\textbf{Base-excitation}}  

\noindent When cantilever is excited from the base and a linear interaction force acts on the tip (Fig. \ref{F-CM_base-tip_excitation_schematic} (a)), following boundary conditions can be applied to solve Eq. \ref{S_spatial_part}:
   \begin{equation*}
         y(0) = A,\qquad \qquad y'(0) = 0
    \end{equation*}
    \begin{equation}
          y''(L) = 0,\qquad EIy'''(L) = (k_i+i\omega\bar{\gamma_i})y(L)
           \label{E-boundary_conditions_base}
    \end{equation}
Where prime ($'$) denotes the derivative of $y(x)$ with respect to $x$. $k_i$ and $\bar{\gamma}_i$ are interaction stiffness and damping coefficients respectively. It is important to note that the interaction force is considered as \textit{linear viscoelastic force}.

Eq. \ref{E-boundary_conditions_base} were solved to get the constants ($a$, $b$, $c$, and $d$) and final solution of Eq. \ref{S_spatial_part}, assuming $z<<1$ and $g<<1$, are as follows (see \cite{rajput2020nano} for detailed derivation): 

\vspace{4mm}

\noindent \textit{\textbf{\small{Slope (bending):}}} 
\vspace{2mm}

       \begin{equation} 
                    \frac{\partial y(x)}{{\partial x}}\Big|_{x=L} = \frac{A}{2k_cL}(-3k_i + \tilde{\rho}\tilde{S} \omega^2 L) - i\frac{A\omega}{2k_cL}(3\bar{\gamma_i} + \gamma_cL)
                    \label{slope_interaction}
        \end{equation} 
\vspace{2mm}
        
Modulus and argument of Eq. \ref{slope_interaction} gives us the amplitude and phase (difference between drive and tip) of bending at $x=L$:

        \begin{equation*}
            R_b =\frac{A}{2k_cL} \sqrt{{(-3k_i+\tilde{\rho}\tilde{S}\omega^2 L)}^2 + \omega^2({3\bar{\gamma_i}+\gamma_cL})^2},
         \end{equation*}    
        \begin{equation}  \label{mag_slope_interaction}
            \theta_b = \arctan{\Big(-\omega \frac{3\bar{\gamma_i}+\gamma_cL}{-3k_i+\tilde{\rho}\tilde{S}\omega^2 L}\Big)} 
        \end{equation}
    
The real and imaginary components of the bending are:

        \begin{equation*} 
            X_b = \frac{A}{2k_c L}(-3k_i + \tilde{\rho}\tilde{S}\omega^2 L),
        \end{equation*}
        \begin{equation}  \label{X_Y_slope_interaction_base}
            Y_b = -\frac{A\omega}{2k_cL}(3\bar{\gamma_i} + \gamma_cL)
        \end{equation}
    
Rearranging the Eq. \ref{X_Y_slope_interaction_base}, the expression for interaction stiffness and dissipation coefficient can be written as:
    
        \begin{equation*}
           k_i = \frac{\tilde{\rho} \tilde{S}\omega^2 L}{3} - \Big( \frac{2k_c L}{3A}  \Big) X_b, 
        \end{equation*}
        \begin{equation}  \label{E-stiff-diss_slope-cont-mass}
           \bar{\gamma_i} = \Big( \frac{2k_c L}{3A\omega}  \Big) Y_b - \frac{\gamma_c L}{3}
        \end{equation}
\vspace{4mm}

\noindent \textit{\textbf{\small{Displacement:}}}
\vspace{2mm}

 \begin{equation}   \label{E-disp_interaction}
            y(L) = \frac{A}{k_c}(k_c - k_i)-i\frac{A\omega}{k_c}\bar{\gamma_i}   
  \end{equation}
  
In Eq. \ref{E-disp_interaction}, the amplitude and phase of cantilever-tip displacement are the modulus and argument of $y(L)$ respectively. These can be written as:

       \begin{equation*} 
            R_{d} =\frac{A}{k_c} \sqrt{(k_c-k_i)^2 + (\omega\bar{\gamma}_i)^2},
        \end{equation*}
        \begin{equation} 
            \theta_{d} = \arctan{\Big(-\omega \frac{\bar{\gamma}_i}{k_c - k_i}\Big)}
        \end{equation}
    
The real and imaginary components of the displacement are:

        \begin{equation*} 
            X_d = \frac{A}{k_c}(k_c - k_i),\qquad 
        \end{equation*}
        \begin{equation}  \label{E-X_Y_disp_interaction}
            Y_d = -\frac{A\omega}{k_c} {\bar{\gamma_i}}
        \end{equation}
        
Eq. \ref{E-X_Y_disp_interaction} can be solved for interaction stiffness and damping coefficient:

    \begin{equation*}
       k_i = k_c \Big(1 - \frac{X_d}{A} \Big), 
     \end{equation*}
     \begin{equation}    \label{E-stiff-diss-cont-mass}
       \bar{\gamma_i} = -\frac{k_c\:Y_d}{A\:\omega}
     \end{equation}
    
Eq. \ref{E-stiff-diss_slope-cont-mass} and \ref{E-stiff-diss-cont-mass} are valid when the stiff cantilever ($g<<1$) is used at low operation frequency ($z<<1$) and small-amplitude ($EIy'''(L) = (k_i+i\omega \bar{\gamma_i}) y(L)$). These criterion must be fulfilled when experiments are performed.

\subsubsection{\textbf{Tip-excitation}} 
\vspace{2mm}

\noindent When the cantilever is excited directly at the tip and also a linear interaction force acting on the tip (Fig. \ref{F-CM_base-tip_excitation_schematic} (b)), following boundary conditions can be applied to solve Eq. \ref{S_spatial_part}:

   \begin{equation*}
         \nonumber 
         y(0) = 0,\qquad \qquad y'(0) = 0, 
     \end{equation*}
    \begin{equation*}  
          y''(L) = 0,\qquad EIy'''(L) = (k_i+i\omega\bar{\gamma_i})y(L)-F_0
    \end{equation*}
   
Where $F_0=k_c  \protect A_0 $ is magnitude of excitation force applied at the tip. $A_0$ is free amplitude of the cantilever end-  when cantilever is far from the surface and interaction force is absent. 

We follow similar protocol to solve Eq. \ref{S_spatial_part} as has been followed in base-excitation section. The final solutions for assumptions $z<<1$ and $g<<1$ are as follows: 
\vspace{4mm}

\noindent \textit{\textbf{\small{Slope (bending):}}} 
\vspace{2mm}

   \begin{equation}
           \frac{\partial y(x)}{{\partial x}}\Big|_{x=L} = \frac{3A_0}{2Lk_c} (k_c - k_i) -i \omega \frac{3A_0}{2Lk_c} \bar{\gamma}_i
            \label{E-slope_tip}
    \end{equation}
    \vspace{2mm}

Modulus and argument of Eq. \ref{E-slope_tip} gives us the amplitude and phase (difference between drive and tip) of bending at $x=L$: 

        \begin{equation*}
            R_b = \frac{3A_0}{2k_c L} \sqrt{(k_c-k_i)^2 + \omega^2 \bar{\gamma_i}^2},
        \end{equation*}
        \begin{equation}       \label{E-mag_slope_interaction}
           \theta_b = \arctan{(-\omega \frac{\bar{\gamma_i}}{k_c-k_i})}
        \end{equation}

Now the real and imaginary components of the bending are
        \begin{equation*} 
            X_b = \frac{3A_0}{2k_c L} (k_c - k_i), 
        \end{equation*}
        \begin{equation}   \label{X_Y_slope_interaction}
            Y_b = - \frac{3A_0 \omega}{2k_c L} \bar{\gamma_i}
        \end{equation}
    
Eq. \ref{X_Y_slope_interaction} can be solved for interaction stiffness and damping coefficient: 
     
         \begin{equation*}  
            k_i = k_c \Big( 1 - \frac{2L}{3A_0} X_b \Big),
         \end{equation*}
          \begin{equation}  \label{E-stiff-diss_CM_tip}
            \bar{\gamma_i} = - \frac{2k_c L}{3A_0 \omega} Y_b
         \end{equation}
    \vspace{4mm}
        
\noindent \textit{\textbf{\small{Displacement:}}}
\vspace{2mm}

    \begin{equation}
            y(L) = \frac{A_0}{k_c} (k_c - k_i) -i \omega \frac{A_0}{k_c} \bar{\gamma}_i
            \label{E-displacement_tip}
    \end{equation}
    
Modulus and argument of Eq. \ref{E-displacement_tip} gives us the amplitude and phase (difference between drive and tip) of displacement at $x=L$: 
    
       \begin{equation*}
            R_d =\frac{A_0}{k_c} \sqrt{(k_c-k_i)^2 + (\omega\bar{\gamma}_i)^2},
       \end{equation*}
        \begin{equation}    \label{mag_disp_interaction}
            \theta_d = \arctan{\Big(-\omega \frac{\bar{\gamma}_i}{k_c - k_i}\Big)}
       \end{equation}
        
The real and imaginary components of the displacement are:
        \begin{equation*} 
            X_d = \frac{A_0}{k_c}(k_c - k_i), 
        \end{equation*}
        \begin{equation}  \label{X_Y_disp_interaction}
            Y_d = -\frac{A_0\omega}{k_c} {\bar{\gamma_i}}
        \end{equation}
        
Eq. \ref{X_Y_disp_interaction} can be solved for interaction stiffness and damping coefficient:

     \begin{equation*}
         k_i = k_c \Big(1 - \frac{X_d}{A_0} \Big),
     \end{equation*}
     \begin{equation}     \label{E-stiff-diss_disp-cont-mass_tip}
       \bar{\gamma_i} = -\frac{k_c\:Y_d}{A_0\:\omega}
     \end{equation}
  
Again Eq. \ref{E-stiff-diss_CM_tip} and \ref{E-stiff-diss_disp-cont-mass_tip} are valid when the stiff cantilever ($g<<1$) is used at low operation frequency ($z<<1$) and small-amplitude ($EIy'''(L) = (k_i+i\omega \bar{\gamma_i}) y(L)$). These criterion must be fulfilled when the experiments are performed. 

It is important to note here that in most AFM measurements, cantilever is base excited and  deflection detection scheme is used  which measures cantilever bending. Eq. \ref{E-stiff-diss_slope-cont-mass}  should be used to analyse the data in such experimental situations \cite{benedetti2016can}. In our previous work we have highlighted how these measurements are prone to artefacts owing to the extraneous  phase contributions. These phase contributions are difficult to account for in  theoretical treatment of cantilever hydrodynamics \cite{rajput2020nano}. It is also noteworthy that  displacement solution for base excited cantilever  (Eq. \ref{E-stiff-diss-cont-mass})  are similar in form to the bending solution of tip excited cantilever (Eq. \ref{E-stiff-diss_CM_tip}) since  $X_d (L) = 2LX_b(L)/3$. The solution for displacement is same for both tip excited or base excited cantilevers (see Eq. \ref{E-stiff-diss-cont-mass} and \ref{E-stiff-diss_disp-cont-mass_tip}) since, for off-resonance operation  $A = A_0$.

\subsection{\textbf{Point-mass model}}

\begin{figure}[hbt!]
  \centering
  \includegraphics[width=1\linewidth]{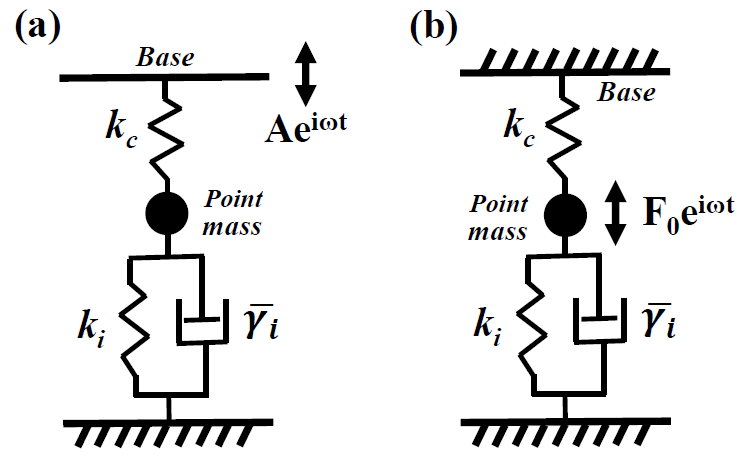}
  \caption {Schematic representation of point-mass approximated cantilever under influence of a linear interaction force. These are representing off-resonance operation where medium damping is negligibly small. (a) Representing the base-excited point-mass approximation where base is driven by $Ae^{i\omega t}$. (b) Representing the tip-excited point-mass approximation where base is rigidly fixed and tip is excited with force $F_0e^{i\omega t}$.}
 \label{F-PM_base-tip_excitation_schematic}
 \end{figure}

In point-mass model, variable separable  method is used  to solve Eq. \ref{E-Euler-Bernoulli-equation} wherein  space and time parts are separately solved. To solve space part, boundary conditions are considered as homogeneous ($y(0) = 0, y'(0) = 0, y''(L) = 0,$ and $y'''(L) = 0$). This provides a set of independent solutions ($y_n(x)$) for a freely oscillating cantilever. Each solution represents an eigen mode. Now solution ($y(x,t)$) is substituted into Eq. \ref{E-Euler-Bernoulli-equation}, multiply with $y_n(x)$ and using the orthogonal property of $y_n(x)$ a set of equations on time-variable is achieved. Each equation for an eigen mode is a  damped harmonic oscillator equation - a linear second order differential equation. Since the fundamental mode makes up for most of  the contribution to  cantilever oscillation, it is considered as the force equation of the freely oscillating cantilever in viscous environment. The excitation force is directly included into the final equation of motion depending on the excitation scheme. The interaction force, which is assumed to be \textit{linear}, is considered as a perturbation to  the forced-damped oscillator and included into the final equation (or in Eq. \ref{E-Euler-Bernoulli-equation} using $\delta$-function). This treatment imposes the assumption of small interaction force compared to the cantilever force (or $k_i << k_c$). Now the final equation of motion of oscillating cantilever under influence of interaction force is a simple forced-damped harmonic oscillator equation with an effective mass ($m^*$), elasticity ($k'$) and damping ($\gamma'$) coefficients. It is important to note that this treatment assumes the cantilever beam as a \textit{point-mass} which is attached to a mass-less spring as shown in Fig. \ref{F-PM_base-tip_excitation_schematic}.  See \cite{garcia2011amplitude} for detailed derivation.

\subsubsection{\textbf{Base-excitation}} 
\vspace{2mm}

\noindent When cantilever is excited from the base sinusoidally $(A e^{i\omega t})$ and tip experiences a \textit{small perturbation} due to a \textit{linear viscoelastic} force $(F_i)$,  it is approximated as a point-mass connected to one end of a spring whose other end is driven (Fig. \ref{F-PM_base-tip_excitation_schematic} (a)). The equation of motion for this point-mass can be written as: 
   \begin{equation}  \label{E-PM equation}
       m^* \frac{\partial^2 z(t)}
        {{\partial t^2 }} + \gamma_c L \frac{\partial z(t)}{\partial t} + k_c (z(t)-A e^{i\omega t}) - F_i = 0
    \end{equation}
Where $m^* = 0.2425\:m_c + m_a$ is effective mass, $m_c$ and $m_a$ are the cantilever mass and hydrodynamic added mass respectively. $F_i = k_i\:z(t) + \bar{\gamma_i}\:\frac{dz(t)}{dt}$ is the interaction force which is assumed to be \textit{linear viscoelastic}. Other parameters have the same meaning as in the continuous-beam model. Eq. \ref{E-PM equation} can be written as:
   \begin{equation}
       m^* \frac{\partial^2 z(t)}
        {{\partial t^2 }} + \gamma' \frac{\partial z(t)}{\partial t} + k' z(t) = k_c A e^{i\omega t}
        \label{E-final point mass equation}
    \end{equation}
Eq. \ref{E-final point mass equation} is an equation of motion for a forced damped harmonic oscillator with effective stiffness $(k' = k_c + k_i)$, effective damping coefficient $(\gamma' = \gamma_c L + \bar{\gamma_i})$, and effective drive force $(k_cA)$. The amplitude ($R$) and phase ($\theta$, phase difference between drive and the oscillator) will be given as:

   \begin{equation*} 
         R = \frac{k_c A}{\sqrt{(k\:' - m^* \omega^2)^2+(\omega \gamma \:')^2}},
    \end{equation*}
     \begin{equation}    \label{E-ampl-phase-point-mass_base}
         \theta = tan^{-1} \Big(\frac{\omega \gamma\:'}{k\:' - m^* \omega^2 }\Big)
    \end{equation}

Further, Eq. \ref{E-ampl-phase-point-mass_base} can be solved for interaction stiffness and damping coefficient for $(\omega << \omega_0)$:
   \begin{equation*}
       k_i = k_c \Big(\frac{A\: cos\theta}{R} - 1\Big),
    \end{equation*}
    \begin{equation}    \label{E-stiff-diss-point-mass_base}
       \gamma\:' = -\frac{k_c\:A\:sin\theta}{R\:\omega}
    \end{equation} 

\subsubsection{\textbf{Tip-excitation}}  
\vspace{2mm}

\noindent When an AFM cantilever is excited from the tip sinusoidally with excitation force $(F_0 e^{i\omega t})$ and tip experiences a \textit{small perturbation} due to a \textit{linear viscoelastic} force $(F_i)$,  it is approximated as a point-mass connected to a massless spring which is driven sinusoidally (Fig. \ref{F-PM_base-tip_excitation_schematic} (b)). The Equation of motion for  this point-mass can be written as: 
   \begin{equation}
       m^* \frac{\partial^2 z(t)}
        {{\partial t^2 }} + \gamma_c L \frac{\partial z(t)}{\partial t} + k_c z(t)-F_0 e^{i\omega t} - F_i = 0
        \label{E-PM_equation_tip}
    \end{equation}
All parameters have the same meaning as described in previous sections. Eq. \ref{E-PM_equation_tip} can be written as:
   \begin{equation}
       m^* \frac{\partial^2 z(t)}
        {{\partial t^2 }} + \gamma' \frac{\partial z(t)}{\partial t} + k' z(t) = F_0 e^{i\omega t}
        \label{E-final_PM_equation_tip}
    \end{equation}
Eq. \ref{E-final_PM_equation_tip} is an equation of motion for a forced damped harmonic oscillator with effective stiffness $(k' = k_c + k_i)$, effective damping coefficient $(\gamma' = \gamma_c L + \bar{\gamma_i})$. The amplitude ($R$) and phase ($\theta$, phase difference between drive and the oscillator) will be given as: 

    \begin{equation*} 
       R = \frac{k_c A_0}{\sqrt{(k\:' - m^* \omega^2)^2+(\omega \gamma \:')^2}},
     \end{equation*}
     \begin{equation}    \label{E-ampl-phase-point-mass_tip}
       \theta = tan^{-1} \Big(\frac{\omega \gamma\:'}{k\:' - m^* \omega^2 }\Big)
     \end{equation}

Further, Eq. \ref{E-ampl-phase-point-mass_tip} can be solved for interaction stiffness and damping coefficient for $(\omega << \omega_0)$:
   \begin{equation*}
       k_i = k_c \Big(\frac{A_0\: cos\theta}{R} - 1\Big),
    \end{equation*}
    \begin{equation}   \label{E-stiff-diss-point-mass_tip}
       \gamma\:' = -\frac{k_c\:A_0\:sin\theta}{R\:\omega}
    \end{equation}
    
 As expected,  for off-resonance conditions  ($A = A_0 $) the solutions for tip excited and the  based excited cantilever is same. Again, the Eq. \ref{E-stiff-diss-point-mass_base} and \ref{E-stiff-diss-point-mass_tip} are valid when stiff cantilever (interaction force ($F_i$) is considered as the perturbation) is used at off-resonance $(\omega << \omega_0)$ and small oscillation amplitude.
 \\

 \begin{center}
     \begin{table}[h!]
     \centering
        \begin{tabular}{|p{1cm}|p{1.5cm}|p{3.6cm}|}
        \hline
        Model & Excitation & Formula\\
        \hline
              &&\\
              & Base & $k_i=\frac{\tilde{\rho}\tilde{S}\omega^2L}{3}-(\frac{2k_cL}{3A})X_b$\\
              &      & $\bar{\gamma_i}=(\frac{2k_c L}{3A\omega})Y_b-\frac{\gamma_c L}{3}$\\ 
              CB$_b$ &&\\
              & Tip     & $k_i=k_c(1-\frac{2L}{3A_0}X_b)$\\
              &         & $\bar{\gamma_i}=-\frac{2k_c L}{3A_0\omega}Y_b$\\
              &&\\
        \hline      
              &&\\
              & Base    & $k_i=k_c(1-\frac{X_d}{A})$\\
              &         & $\bar{\gamma_i}=-\frac{k_cY_d}{A\omega}$\\
              CB$_d$ &&\\
              & Tip      & $k_i=k_c(1-\frac{X_d}{A_0})$\\
              &          & $\bar{\gamma_i}=-\frac{k_cY_d}{A_0\omega}$\\      
              &&\\
        \hline
              &&\\
              & Base   & $k_i=k_c(\frac{A cos\theta}{R} - 1)$\\
              &        & $\gamma'=-\frac{k_c A sin\theta}{R\omega}$\\
              PM &&\\
              & Tip    & $k_i = k_c(\frac{A_0\cos\theta}{R} - 1)$\\
              &        & $\gamma'=-\frac{k_c A_0\:sin\theta}{R\:\omega}$\\
              &&\\     
         \hline
        \end{tabular}
        \caption{\label{tab:formulas}For the CB model, $X_b$, $Y_b$ and $X_d$, $Y_d$ are the $X$ and $Y$ components of bending and displacement signals respectively. In the PM model, A and A$_0$ are the drive amplitudes of the base and tip respectively. CB$_d$ and CB$_b$ stands for displacement and bending solutions of continuous beam model.}
    \end{table}
\end{center}

\section{Experimental}
\subsection{Atomic Force Microscopy}

\begin{figure}[hbt!]
\includegraphics[width=1\linewidth]{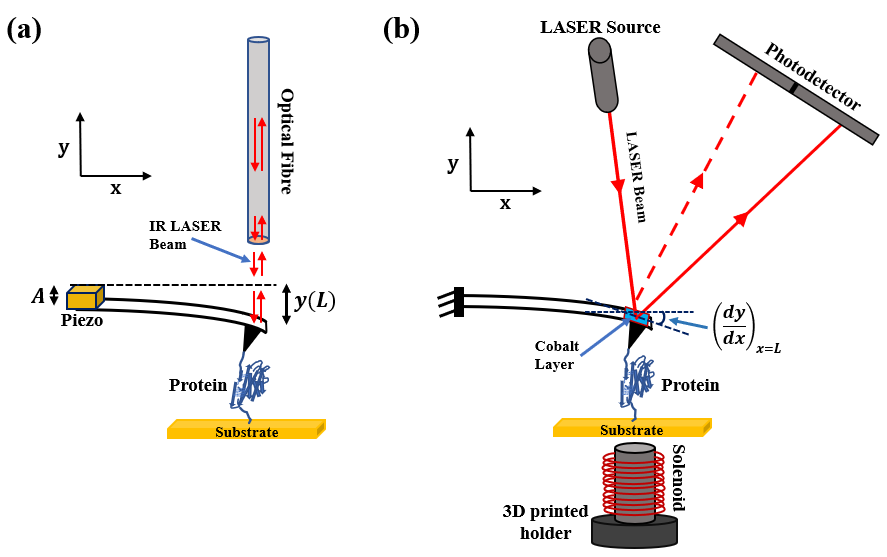}
\caption{(a) Schematic of interferometer based  AFM equipped with base-excitation mechanism. Cantilever is attached with a piezoelectric block which excites its base. Interferometric-detection  scheme measures displacement ($y(t)$) of cantilever end. (b) Representing tip-excitation setup equipped with a deflection-detection mechanism. A thin layer of cobalt is coated at the tip-end, which is used to excite cantilever's tip using an electromagnet placed beneath it,  base of the cantilever is rigidly fixed. Optical-deflection-detection scheme measures the bending ($dy(t)/dx$) of the cantilever end. A protein with one domain is tethered between the substrate and cantilever-tip.}
\label{fig:F-slope_and_displacement_detection}
\end{figure}

We performed two types of AFM measurements, i) Base excited cantilever together with interferometer based detection scheme which, as shown in Fig. \ref{fig:F-slope_and_displacement_detection} (a),  measures cantilever displacement.
The interferometer is home-built with five-axis inertial slider allowing nanopositioning of fiber on the back of the cantilever. It is further aligned exactly perpendicular to the length of the cantilever,  so that the end of the fiber and the backside of cantilever form a fabry-perot etalon. This design enhances the sensitivity allowing a thermally limited measurement of the cantilever's displacement \cite{patil2005highly}. Next, we performed experiments with ii) tip excited cantilever with deflection detection scheme which, as shown in Fig. \ref{fig:F-slope_and_displacement_detection} (b),  measures cantilever bending at the tip end. The deflection detection type measurement scheme is available in all commercial AFMs and it is preferred due to its relative ease of operation. The deflection detection set-up together with base-excitation is more prone to artefacts in the measurement and quantification of visoelastic parameters of nanoscale entity under investigation is difficult \cite{raman2008cantilever,rajput2020nano}. Therefore,  we avoided exciting the cantilever from base in this case.  It is excited at the tip-end. Such excitation requires a magnetic spot  at the back of cantilever. This is achieved by coating a  spot of Cobalt having  $\sim$ 30 $nm$ thickness on the back of the cantilever, near the tip. A home-built solenoid coil is used to excite the tip sinusoidally. It ensures the tip is preferentially excited compared to the other parts of the cantilever and surrounding liquid.

\subsection{Sample preparation}

The human cardiac muscle protein I27$_8$ gene was cloned in the pET-23a vector. I27$_8$ was transformed in BL21(DE3) and induced at 0.6 OD with 1mM IPTG for 6 hours at 37 °C. Cells were pelleted and stored at -40°C. For purification, the frozen bacterial pellet was resuspended in 1XPBS $pH$ 7.4 supplemented with protease inhibitor cocktail (Roche). After resuspension, the cells were lysed by sonication in an ice water bath. The lysate was spun down at 18,500 $g$ for 30 $minutes$ and the supernatant was incubated with His-Pur cobalt resin (Thermo Scientific) for 1 $hour$ at 4°C. The supernatant was then poured into the PD-10 column, and the resin was washed with 150 $ml$ of 1XPBS $pH$ 7.4 to get rid of non-specifically bound proteins. Protein was eluted using 1XPBS with 250 $mM$ imidazole. Protein purity was checked using coomassie gel. Purified protein fractions were pooled and dialyzed overnight against 1XPBS $pH$ 7.4 buffer to remove imidazole. For short-term storage (about a week), proteins were kept on ice at 4°C. For long-term storage (about 3 $months$) the protein was flash-frozen in liquid nitrogen with the addition of 10$\%$ glycerol. Frozen protein was dialyzed overnight against 1XPBS at 4°C to remove glycerol.

Sample is prepared in manner similar to all protein pulling experiments using AFM \cite{rajput2020nano}. A 100 $\mu l$ protein sample in PBS buffer ($pH$ 7.4) with 10 $mg \: ml^{-1}$ was drop casted on a freshly evaporated gold coated cover-slip. It is allowed to incubate for 30 minutes along with the cleaning with fresh buffer two-to-three times. The liquid cell is filled with $\sim$ 600 $\mu l$ buffer and used  for the measurement.

\subsection{Cantilevers}
Rectangular cantilevers from  Micromasch with resonance frequency in the range of 7-15 $kHz$ in water,    with stiffness  0.4 - 1 $N/m$ and dimensions - length $= 250 - 300 \:\mu m$, width $= 35\: \mu m,$ thickness $= 1 \: \mu m$- are used for experiments. The cantilevers are calibrated before each measurement to determine their stiffness using thermal tuning method \cite{butt1995calculation}.

To facilitate  excitation from the tip end,  a thin layer of Cobalt is coated on the back of the cantilever near the tip end. 
For this, the entire cantilever, except the end position, was carefully masked using a metal sheet. Using thermal evaporation deposition method  chromium-cobalt were coated on the masked cantilever. The metal sheet was then removed. This leads to only the Cr-Co film remaining at the end of the cantilever. It was then sputter-coated with gold from all sides to protect it from the buffer in which the experiments were performed.

\subsection{Unfolding experiments}
In commercial AFM, the oscillating cantilever is approached using a servo control towards the surface on which the octamers of I27$_8$ are sparsely coated. Once engaged, the tip is held at a fixed load of few $nN$ on the surface. After waiting for few seconds, the cantilever is pulled back. A digital lock-in amplifier (SRS830, Stanford Research Systems, California, US) was used to get the amplitude, phase, in phase $X$-component and out-of-phase $Y$-component during the measurement. We collect 256 force curves on a grid of 10 $\mu m$ X 10 $\mu m$ in this way. As seen with other attempts to measure force extension curves of protein unfolding, we see 5$\%$ of these curves carrying a finger-print pattern of an octamer unfolding. The amplitude and phase along with $X$ and $Y$ components are recorded using a data acquisition card. The curves showing  more than 6 unfolding events are further analysed.  Fig. \ref{fig:F-raw_data_SAAFM_XYAP_133Hz} and  \ref{fig:F-raw_XYAP_mag_133Hz} show  representative curves of various parameters of the oscillating cantilever versus the extension as the cantilever is pulled away from the surface.

In home-built setup, sample-cell is approached to the oscillating cantilever using an amplitude feedback control. Once approached at desired set-point amplitude, sample-cell is moved towards the cantilever to allow the tip to be in contact with the sample surface. Tip is allowed to be in contact with the surface for few seconds and then sample is retracted back.  The amplitude and phase versus extension curves were collected  on a grid of 3 $\mu m$ X 3 $\mu m$. LabVIEW programs are used as interface to control the instrument and data acquisition.

\section{Results}
The sequential unfolding of  protein octamer (I27$_8$)  can be used to test the validity of point mass model in AFM experiments  to quantitatively extract viscoelastic properties of nanoscale systems. Such unfolding of  (I27$_8$) is routinely reported in literature using static AFM.  We  record parameters of oscillating cantilever  while octamer is  unfolded as  described in section 3.4.  The data is  analysed using both CB and PM model to compare  results with each other. 

\subsection{Displacement-detection, Base-excited cantilever}

\begin{figure}[hbt!]
\includegraphics[width=1\linewidth]{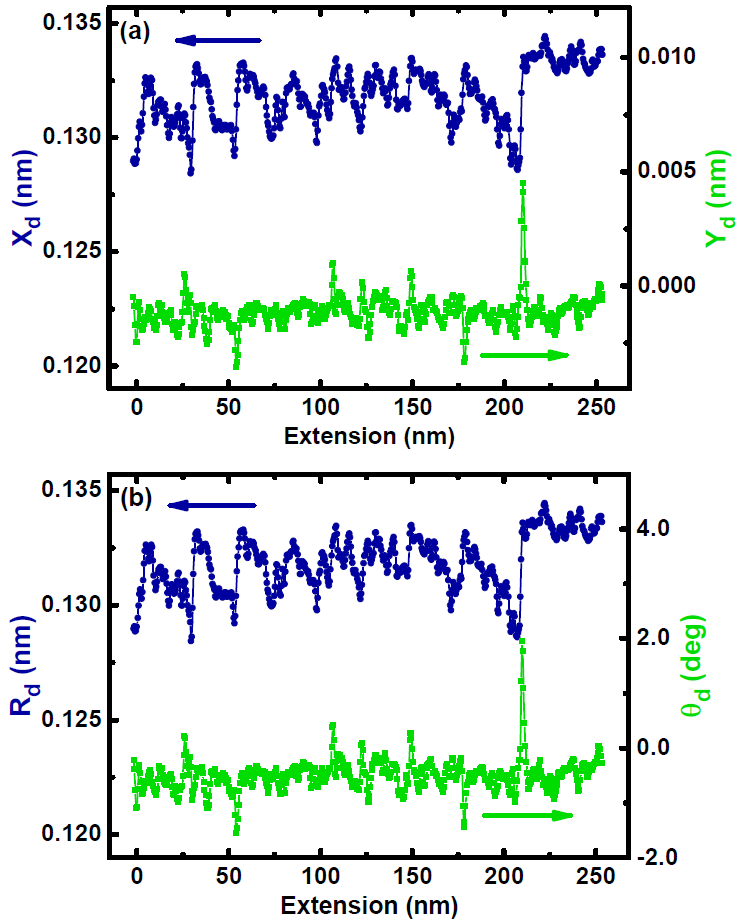}
\caption{Base-excited unfolding experiment was performed on polyprotein using interferometer based detection scheme wherein cantilever displacement is measured.  (a) $X$-component (solid circle in dark blue) and $Y$-component (solid square in green) of displacement. (b) Amplitude (solid circle in dark blue) and $Phase$ (solid square in green) of displacement.}
\label{fig:F-raw_data_SAAFM_XYAP_133Hz}
\end{figure}

Using a home-built, fibre-interferometer based AFM \cite{patil2005highly}    measurements are performed at low pulling speed ($\sim 15$ $nm/s$). Fig. \ref{fig:F-raw_data_SAAFM_XYAP_133Hz} (a) shows as collected raw data of the in-phase $X$-component and out-of-phase $Y$-component of the cantilever displacement measured using the interferometer. Fig. \ref{fig:F-raw_data_SAAFM_XYAP_133Hz} (b)  shows amplitude and phase of the displacement. 

While performing these measurements care is taken so that the approximations used to derive Eq. \ref{E-stiff-diss-cont-mass} and \ref{E-stiff-diss-point-mass_base} are valid. To satisfy the condition $\omega << \omega_0$, experiments are performed at low frequency ($f$ = 133 $Hz$), whereas the fundamental frequency ($f_0$) of the cantilever in liquid is 12.7 $kHz$. At true off-resonance regime, the amplitude and phase response with frequency were flat and phase was $\sim 0 \: deg$ \cite{rajput2020nano}, which confirms that the variation in amplitude and phase will occur only due to interaction stiffness and damping forces respectively and not due to resonance frequency shift \cite{hoffmann2001direct, patil2005small}. The off-resonance operation avoids issues that may  arise at resonance operation due to cantilever's complex amplitude and phase response \cite{hoffmann2001direct, jeffery2004direct, hoffmann2001energy, khan2010dynamic}. 
The stiffness of unfolded protein chain is reported to be around 20 $mN/m$ \cite{rajput2020nano}. We used cantilever stiffness $k_c$ = 0.69 $N/m$, so that $k_i << k_c$. To get an accurate estimate of stiffness  from these linear approximation based models, the experiments must be performed in the linear interaction force regime. This can be achieved  by keeping the oscillation amplitude of the cantilever-tip below  a certain value so that the potential can be approximated as linear over the entire oscillation amplitude.  The oscillation amplitude for the data shown in Fig. \ref{fig:F-raw_data_SAAFM_XYAP_133Hz} were $\sim$ 0.13 $nm$,  which is below the persistence length ($\sim$ 0.4 $nm$) of this protein. 

When experiments were performed on single I27$_8$ molecule with above discussed three criterion, the change in amplitude due to molecule stretching is very small ($\sim$ 0.01 $nm$). To detect such a small signal we need a high sensitivity as well as a good signal to noise ratio. In our home-built, interferometer based AFM, we routinely get $\sim$ 40 $mV/$\AA \space sensitivity and $\sim 5\: pm$ noise-floor at the measurement bandwidth which allows us to detect such a small signal.

\subsection{Deflection-detection, tip-excited cantilever}

\begin{figure}[hbt!]
\includegraphics[width=1\linewidth]{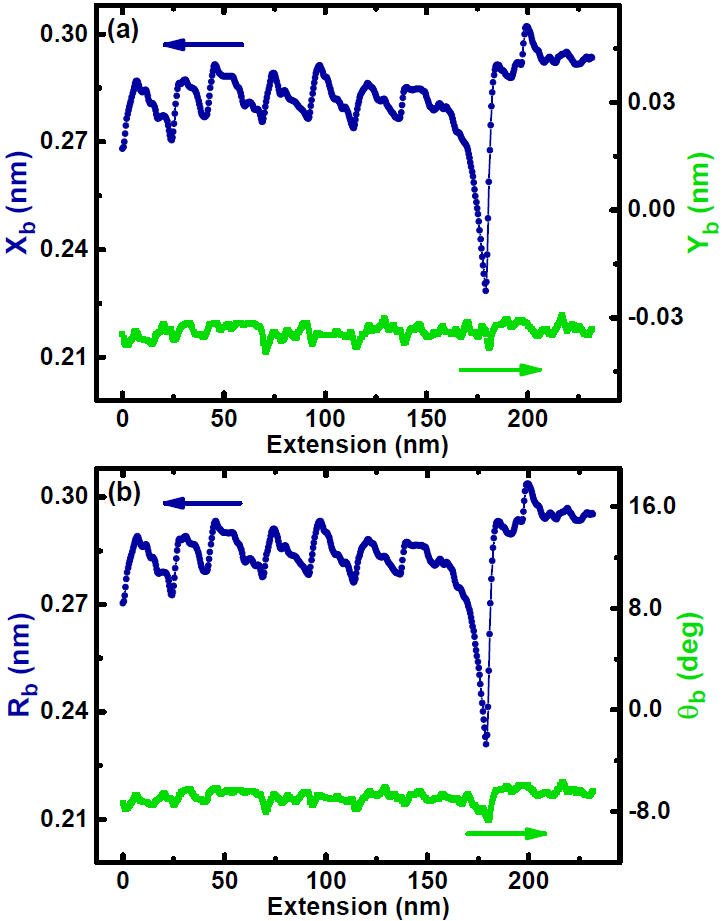}
\caption{Tip-excited unfolding experiment was performed on polyprotein in deflection  detection scheme wherein cantilever bending is measured. (a) $X$-component (solid circle in dark blue) and $Y$-component (solid square in green) of bending. (b) Amplitude (solid circle in dark blue) and $Phase$ (solid square in green) of bending.}
\label{fig:F-raw_XYAP_mag_133Hz}
\end{figure}

We performed measurements using deflection detection scheme.  For these measurements we used magnetic excitation to drive the cantilever tip. Cantilever stiffness was 0.41 $N/m$ and resonance frequency was 7.5 $kHz$ in water. The cantilever is coated with thin layer  of cobalt on the backside of the cantilever near the tip-position as described in section 3.1.     
The amplitude or phase versus extension are recorded in a manner similar to interferometer based measurements. Fig. \ref{fig:F-raw_XYAP_mag_133Hz} shows the raw data of unfolding I27$_8$ molecule performed in deflection-detection based AFM using the tip-excitation method. Once again care was taken so that the approximations used in  the mathematical models are valid.   Experiments were performed at truly-off resonance regime ($f = 133 \: Hz$) and the free amplitude was $\sim$ 0.29 $nm$.

\subsection{Data analysis} 
The analysis is performed using both  CB and PM model on all the data sets. It is worthwhile to mention here that both models describe  hydrodynamics of a cantilever tethered with a protein. We do not fit experimental data to models from polymer physics, which describe  entropic elasticity of single polymer chains such as unfolded proteins.    
In Fig.  \ref{fig:F-raw_data_SAAFM_XYAP_133Hz} and \ref{fig:F-raw_XYAP_mag_133Hz}, it can be observed that only $X$-component and the amplitude  shows variation when the domain was stretched, whereas the $Y$-component  and phase is featureless. A quick glance at the results from both CB and PM model suggests that it indicates immeasurably low  dissipation in the unfolded protein chain.  This is consistent with our previous  observation\cite{rajput2020nano}. Therefore, we used the amplitude and in-phase $X$-component for further analysis to estimate stiffness.

\begin{figure}[hbt!]
\includegraphics[width=1\linewidth]{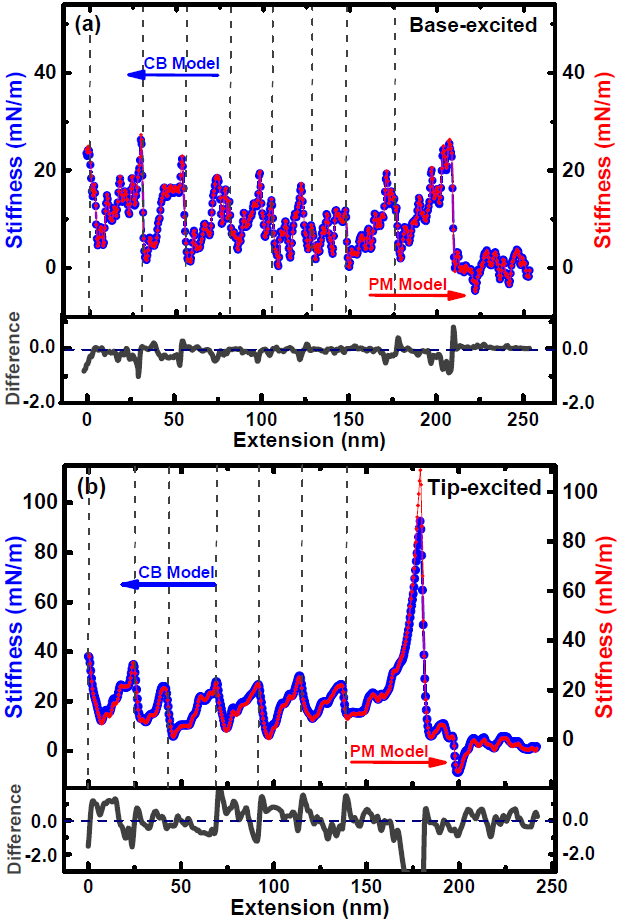}
\caption{Stiffness calculated using CB model (solid circle in blue color) and PM model (Solid diamond in red color). (a) Represents base-excited displacement measurement. (b) Represents tip-excited bending measurement. Difference between calculated stiffness using CB and PM model were plotted at the bottom of each figure.}
\label{fig:F-stiffness_CM_PM_base_&_tip} 
\end{figure}

The data in Fig. \ref{fig:F-raw_data_SAAFM_XYAP_133Hz}(a) and (b) were analyzed using CB model (Eq. \ref{E-stiff-diss-cont-mass}) and PM model (Eq. \ref{E-stiff-diss-point-mass_base}) and the stiffness-extension profile  is  plotted in Fig. \ref{fig:F-stiffness_CM_PM_base_&_tip} (a). The solid-circle in blue represents the stiffness estimated using CB model and solid-diamond in red estimated using PM model. In short, the in-phase component of displacement $X_d$ versus extension is converted to stiffness-extension using Eq. \ref{E-stiff-diss-cont-mass}. The values of $k_c$ and $A$ are 0.69 $N/m$  and 0.13 $nm$ respectively. The observed amplitude-extension curve is converted to stiffness-extension curve using Eq. \ref{E-stiff-diss-point-mass_base}. We observed that analysis using both models yield same stiffness for the unfolded I27$_8$ molecule. It is in the range of 0 to $\sim$ 20 $mN/m$. We obtained consistent results when the measurement was repeated for a range of amplitudes- 0.05 $nm$ \space to 0.3 $nm$, and frequencies- 100 $Hz$ to 1000 $Hz$ (Data not shown).

Similarly,  the data in  Fig. \ref{fig:F-raw_XYAP_mag_133Hz} (a) is  analyzed using CB model. Eq. \ref{E-stiff-diss_CM_tip} is used to convert the in-phase bending component, $X_b$, versus extension data to obtain stiffness-extension profile.  Eq. \ref{E-stiff-diss-point-mass_tip} is used  to calculate the stiffness-extension profile from amplitude-extension curves shown in Fig. \ref{fig:F-raw_XYAP_mag_133Hz} (b). The value of $A_0$ is 0.29 $nm$, $k_c$ is 0.41 $N/m$. These values are experimentally determined. It is  plotted in Fig. \ref{fig:F-stiffness_CM_PM_base_&_tip} (b). The solid-circle in blue represents the stiffness estimated using CB model and solid-diamond in red estimated using PM model. It can be observed that analysis using both models  yield  same stiffness profiles both for the tip excited as well as base excited cantilevers. The experimental parameters for both measurements are listed in Table \ref{tab:experimental_parameters}. 

Fig \ref{fig:F-stiffness_CM_PM_base_&_tip} shows the stiffness calculated using CM and PM models along with the difference in the corresponding values for both measurements. The difference in the stiffness values from both models are $\sim$ 1 $mN/m$. We computed the errors in estimating the stiffness for both methods. For this analysis we took $5\%$ error in stiffness calibration of the cantilevers by thermal tuning. The errors in $X$, $Y$ components and  amplitudes are estimated by taking a standard error in these quantities when the cantilever is far from the substrate and is not tethered with the molecule. These errors are 2.9 $mN/m$ and 2.6 $mN/m$ for interferometer-based measurements and deflection detection measurements respectively. Both these error estimates are considerably higher compared to the difference between the two models, indicating that both models yield  same values of stiffness within the experimental uncertainties.

I27$_8$ is a multi-domain protein with 8 domains, each with  contour length of 25 $nm$.  The stiffness of poly-peptide chain after unfolding is much lower than the folded domains and hence our measurements in Fig. \ref{fig:F-stiffness_CM_PM_base_&_tip} measure stiffness of the unfolded chain. This stiffness is entropic  arising out of chain taking very few out of myriad  possible configurations when not under force. The entropic stiffness increases as the chain is stretched further. The dotted vertical lines in Fig. \ref{fig:F-stiffness_CM_PM_base_&_tip} are $\sim$ 25 $nm$ apart from each other. As the stiffness of the chain reaches a critical value, and through it  cantilever is able to exert  enough force on  folded domains, one of them unfolds. This produces an unfolding event and the process continues till all the domains are sequentially unfolded. Similar to static pulling experiments, the dynamic stiffness measurements also show a finger-print pattern of unfolding events as seen in Fig. \ref{fig:F-raw_data_SAAFM_XYAP_133Hz} and \ref{fig:F-raw_XYAP_mag_133Hz}. The final detachment of the protein from either tip or the substrate occurs at around 200 $nm$ in Fig. \ref{fig:F-raw_data_SAAFM_XYAP_133Hz}. It occurs around 175 $nm$  in Fig. \ref{fig:F-raw_XYAP_mag_133Hz}. This difference is  because of 8 unfolding events compared to 7. We have used  cantilevers with different stiffness  in both experiments. The agreement between CB and PM model is seen to be consistent for both these cantilevers.

\begin{center}
\begin{table}
\begin{tabular}{ | m{3.7cm} | m{0.9cm} | m{0.9cm} | m{0.9cm} | } 
\hline
Experiment & $k_c$ ($N/m$) & $\omega /  2\pi$ ($Hz$) & A, A$_0$ ($nm$) \\ 
\hline
Displacement-detection, Base-excited & 0.69 & 133 & 0.13 \\ 
\hline
Deflection-detection, tip-excited & 0.41 & 133 & 0.29 \\ 
\hline
\end{tabular}
\caption{\label{tab:experimental_parameters}The experimental parameters used to calculate stiffness (shown in Fig. \ref{fig:F-stiffness_CM_PM_base_&_tip}) from observable parameters (shown in Fig. \ref{fig:F-raw_data_SAAFM_XYAP_133Hz} and \ref{fig:F-raw_XYAP_mag_133Hz}) using Eq. \ref{E-stiff-diss-cont-mass}, \ref{E-stiff-diss_CM_tip}, \ref{E-stiff-diss-point-mass_base}, and \ref{E-stiff-diss-point-mass_tip}.}
\end{table}
\end{center}

\section{Discussion}

Stiffness of an unfolded protein chain under force has important implications in the protein folding dynamics and determines the rate of collapse from chain to a molten globule.  We discuss our results obtained through Fig. \ref{fig:F-raw_data_SAAFM_XYAP_133Hz} and \ref{fig:F-raw_XYAP_mag_133Hz} in the context of other attempts to measure stiffness of I27$_8$ using off-resonance dynamic AFM measurements. Khatri et al. used a force clamp measurements to deduce the stiffness of I27$_8$ by measuring the thermal noise in the cantilever. They report $k_i L_i$ = 1000-2000 $pN$ \cite{khatri2008internal}. Taking the length of the stretched molecule, $L_i$  to be 25 $nm$, the stiffness of the unfolded chain goes from 0 to 40-80 $mN/m$. Benedetti et al. used off-resonance dynamic AFM using base-excited cantilever and deflection detection scheme. They estimated the stiffness range of unfolded chain to be 4.0 $\pm$ 0.15 $mN/m$. Both these values are not consistent with the derivative of  static force extension measurements. Our measurements with interferometer based AFM as well as deflection detection AFM yield a consistent value of stiffness estimated from two different models for data analysis. It changes from 0 to $\sim$ 20 $mN/m$ as individual domains are unfolded and agrees well with  static measurements. Once again the agreement of static and dynamic stiffness, as also shown in our previous work \cite{rajput2020nano}, suggest that the damping coefficient must be immeasurably low since these measurements are done at entirely different strain rates (a factor of $10$ higher). It means that no irreversible processes contribute to unfolding of the molecule under force.  This is also emphasised by Liu et al. in their measurement of single macromolecules \cite{liu1999relationship}. The conspicuous absence of dissipation signal has another important implication. Many mechanical measurements on single proteins  have suggested that the effective  diffusion coefficient of an unfolded chain, determined by its  damping coefficient, is remarkably low ($\sim 10^3 \: nm^2/s$)  compared to molecular dynamics simulations or optical techniques such as Fluorescence Resonance Energy Transfer  ($\sim 10^8 \: nm^2/s$). Berkovich et al. have argued that  mechanical measurements do not probe the damping coefficient of the unfolded chain, but instead that of the probe to which the molecule is tethered to \cite{berkovich2012rate}. The  non-observation of dissipation in both types of measurements presented here  supports the argument of Berkovich et al.

Since both PM  and CB model yield same results on  experimental data collected using two types of measurement schemes, it is important to ask the question- How does a model, which accommodates geometric details of  the cantilever into its dynamics fares similar to a simplistic point-mass approximation? It must be noted that the interaction force in both models is treated like a perturbation. While using the CB model to derive  expressions relating  stiffness and damping to experimentally measured  quantities such as $X$ and $Y$ components (Eq. \ref{E-stiff-diss-cont-mass} and \ref{E-stiff-diss_CM_tip}), it is assumed that the interaction stiffness is much smaller than the cantilever stiffness ($g<< 1$). This implies that the relative change in amplitude due to interaction is minimal (< 10 $\%$). The PM model also works with same assumption.  The interaction force can be considered as a perturbation on the oscillating cantilever when the entire dynamics is dominated  by  forces other than the one acting at the tip-end alone. However, when the interaction force is  comparable to these forces, the effects of altered boundary condition will be significant. It is important to note that in such a scenario, Eq. \ref{E-stiff-diss-cont-mass} and \ref{E-stiff-diss_CM_tip} will have additional terms since the condition $g<<1$  is not satisfied. This assumption is crucial while working with both models and is also the reason behind both of  them yielding same results on off-resonance measurements.  

To observe the difference in stiffness estimates using two models at high interaction force in experiment, we analyzed the data when the tip is in strong interaction with substrate resulting in high stiffness compared to stretching of molecules. It is seen  that the stiffness estimated using both models coincides up to $k_i \approx 0.1 k_c$. Beyond this, the stiffness estimated from PM starts deviating from CB (see supplementary Fig. S1). It has already been stated that beyond $k_i \approx 0.1 k_c$ limit, the stiffness estimates using  both models will not be reliable. This is because the assumption of $k_i << k_c$ in derivation of both models is not valid anymore. Secondly, the measurements have to be truly off resonance. As pointed out in \cite{rajput2020nano}, this limit is below 1 $kHz$ for cantilevers having resonance around 15 $kHz$. In order to use PM model as well as the condition of linearity is also important, namely the oscillation amplitude has to be small enough so that the potential the tip is under can  linearized over the entire amplitude. This is the parameter space in which PM model can be used for data analysis of off-resonance dynamic AFM data.

We compare our results  with other attempts to measure stiffness of single molecules using dynamic AFM. There are many reports in the literature about measurement of dissipation in single polymer chains. They have used PM model to analyse their dynamic AFM data.  Recently, Benedetti et al. \cite{benedetti2016can}  have proposed a method to estimate viscoelasticity of single proteins using CB model (Eq. \ref{E-stiff-diss_slope-cont-mass}). In our previous work, we have used the same method to analyse our based-excited,  deflection detection AFM measurements. It is seen in these experiments that the phase signal contains features as the poly-protein unfolds sequentially, whereas the out-of-phase $Y$-component of the amplitude is featureless. The use of PM model in such measurements,  wherein phase is interpreted as dissipation, certainly produces artefacts. Many groups have used this strategy to report observation of dissipation in single molecules \cite{liang2019investigating, kawakami2006viscoelastic, bippes2006direct}. The phase change actually occurs  due to entropic stiffness increase  as the unfolded protein is stretched further. Our  work has additionally highlighted the effect of extraneous phase contributing to artefacts even to the $Y$-component when deflection detection scheme is used to measure the cantilever bending at the tip-end. There are many sources of  extraneous phase contributions  and they are almost intractable in  experiments. This clearly indicates that it is extremely difficult to perform based-excited, deflection detection AFM measurements to estimate error-free viscoelastic response of nano-scale entities in liquid environments. 

It is important to note that in a typical commercial AFM the off-resonance measurements are difficult to perform. The cantilever bending at the tip end  is negligibly small compared the displacement when the base is excited \cite{garcia2011amplitude, rajput2020nano, patil2005highly}. In order to perform true off-resonance measurements with based excited cantilevers, one needs to measure displacement which is only possible with interferometer detection schemes set-up for this purpose. As shown in this work, the tip excitation coupled with deflection detection is another strategy to perform true off-resonance AFM measurements with simple PM model to analyse the data. 

Furthermore, one may  not get such an  agreement between application of  CB and PM models on  base-excited experiments performed with deflection-detection based AFMs,  as the base amplitude is not directly  measured. One has to rely on indirect ways to estimate this parameter \cite{jai2007analytical, de2010dissipation}. This may introduce significant errors in  final results. One can use the deflection detection based AFMs to perform small-amplitude and off-resonance measurements reliably when the cantilever-tip is directly excited. It can be done in various ways such as magnetic  \cite{o1994atomic} or  thermal excitation \cite{fukuma2009wideband, PhysRevB.81.054302}.

In Fig. \ref{fig:F-raw_data_SAAFM_XYAP_133Hz}, $Y$ and $\theta$ signals do not show any change while the  molecule stretched. This has already been reported earlier and it has been concluded that the dissipation in single molecule is below the detection limit of AFM \cite{rajput2020nano}. On the other hand, the dissipation in the confined molecular layers of liquid is  observed in the past with clear features in the phase signal\cite{rajput2020nano,khan2010dynamic, patil2006solid}.  The dissipation of confined molecular layers analyzed using both CB and PM models yield  the same profiles (data not shown). This implies that PM model is adequate to predict accurate viscoelasticity of nano-scale systems if care is taken while performing the experiments. The measurements performed using small amplitude interferometer-based AFM which are analysed with PM model on nanoconfined molecular layers \cite{patil2006solid, khan2010dynamic} are accurate estimates of its viscoelasticity.

\section{Conclusion}
We have performed base-excited dynamic atomic force microscopy on a single titin I27$_8$ molecule using a fiber-interferometer based home-built AFM as well as using an conventional AFM equipped with elaborate tip-excitation scheme. In our experiments, we strictly adhere to fulfilling three important criterion so that approximations used to obtain meaningful expressions using both CB and PM models are valid. These criterion are  1) truly off-resonance operation, 2) small oscillation amplitudes so that the measurements are linear, and 3) cantilevers with many orders of magnitude stiffer than the interaction stiffness. It is shown that when care is taken to fulfill these criterion while performing experiments, both the continuous-beam (CB) and the point-mass (PM)  treatment yield same results. It also confirms the past estimates of stiffness and dissipation of molecular layers of liquids using interferometer based AFM are accurate.
 In conclusion, the PM model is adequate to explain the cantilever-molecule system and can be reliably used to extract the viscoelasticity of nano-scale interactions given that experiments are performed at truly off-resonance regime and all the assumptions taken in modeling the cantilever dynamics are fulfilled in the experiments.

\section*{Acknowledgments}
Authors would like to acknowledge Prof. A.S.R. Koti (TIFR, Mumbai) for providing the plasmid, Prof Thomas Pucadyil and Prof. Amrita Hazra (IISER, Pune) for providing the facility for protein extraction and purification. Authors would like to acknowledge Saurabh Talele- he played a crucial role in building the displacement-detection setup. SR, SD acknowledge fellowship from IISER Pune. VJA acknowledges fellowship from INSPIRE-DST. SCK acknowledges fellowship from CSIR. The work is carried out using the grant from the Wellcome-Trust-DBT India alliance through  intermediate  fellowship to SP (500172/Z/09/Z). \newline

\bibliographystyle{unsrt}   
\bibliography{referances}

\end{document}